\documentclass[reprint,aps,prb,floatfix,superscriptaddress,showpacs]{revtex4-1}\usepackage[utf8]{inputenc}
\usepackage{mathtools}
\usepackage{amssymb}
\usepackage{graphicx}
\usepackage{hyperref}
\setcitestyle{super} 

\hypersetup{
  colorlinks   = true, %Colours links instead of ugly boxes
  urlcolor     = blue, %Colour for external hyperlinks
  linkcolor    = red, %Colour of internal links
  citecolor   = green %Colour of citations
}

\begin{document}

\title{Effects of scarring on quantum chaos in disordered quantum wells}% Force line breaks with \\

\author{J. Keski-Rahkonen}
\affiliation{Computational Physics Laboratory, Tampere University, 33720 Tampere, Finland}
\author{P. J. J. Luukko}
\affiliation{Computational Physics Laboratory, Tampere University, 33720 Tampere, Finland}
\author{S. Åberg}
\affiliation{Mathematical Physics, Lund University, 22100 Lund, Sweden}
\author{E. R{\"a}s{\"a}nen}
\affiliation{Computational Physics Laboratory, Tampere University, 33720 Tampere, Finland}

\date{\today}

\begin{abstract}
The suppression of chaos in quantum reality is evident in quantum scars, i.e., in enhanced probability 
densities along classical periodic orbits. They provide opportunities in controlling  quantum transport in nanoscale quantum systems. 
Here, we study energy level statistics of perturbed two-dimensional quantum systems exhibiting 
recently discovered, strong perturbation-induced quantum scarring. In particular, we focus on the 
effect of local perturbations and an external magnetic field on both the eigenvalue statistics and scarring. 
Energy spectra are analyzed to investigate the chaoticity of the 
quantum system in the context of the Bohigas-Giannoni-Schmidt conjecture. We find that in systems where strong perturbation-induced scars are present, the eigenvalue statistics are mostly mixed, i.e., between Wigner-Dyson and Poisson pictures in random matrix theory. However, we report interesting sensitivity of both the eigenvalue statistics to the perturbation strength, and analyze the physical mechanisms behind this effect.
\end{abstract}

\maketitle

%\tableofcontents

\section{\label{sec:level1}Introduction }

Chaotic behavior is ubiquitous and plays an important part in most fields of science.  However, reconciling quantum formalism with classical physics has been a long-standing challenge. The fundamental disconnection poses a challenge to quantum-classical correspondence~\cite{chaology}, and has motivated a long-standing search for quantum signatures of classical chaos.~\cite{Gutzwiller,Stockmann,Haake, Eckhardt}

On the classical side, it is widely acknowledged that generic classical systems are difficult to predict due to chaos. However, on the quantum side, we can push the limit further by utilizing quantum coherence for our benefit. A \emph{quantum scar} is a striking visual example of quantum mechanical suppression of chaos: a track of enhanced probability density
in the eigenstates of a quantum system that occur along short unstable periodic orbits (POs) of the chaotic classical counterpart.~\cite{Heller, Kaplan}

In addition to the conventional scars, a new type of quantum scarring was recently discovered~\cite{Luukko} in a two-dimensional (2D), radially symmetric quantum well disturbed by local perturbations (Gaussian ``bumps'' in the potential). Later, similar scars were observed in a perturbed 2D harmonic oscillator exposed to an external magnetic field.~\cite{controllability} Some of the high-energy eigenstates are scarred exceptionally strongly by short POs of the corresponding \emph{unperturbed} system. These scars have a similar appearance to the conventional scars, but a fundamentally different origin: they stem from the classical resonances in the unperturbed system resulting in semiclassical near-degeneracies (resonant sets) in the unperturbed quantum system. Localized bumps then generate scarred eigenstates out of the resonant sets as these perturbed-introduced (PI) scars effectively extermize the perturbation.

Before the concept of quantum scarring, it was naively assumed that the quantum states of a classically chaotic system would be featureless and random, since the classical system fills the available phase space evenly with preserved total energy, up to random fluctuations. In this chaotic regime, any control in quantum transport would be tedious to realize. However, it has been shown~\cite{Luukko} that PI scars can be exploited to propagate quantum wave packets in the system with very high fidelity. Furthermore, the existence, geometry, and orientation of PI scars are controllable as demonstrated in Ref.~\onlinecite{controllability}. Thus, PI scars may indicate a path to coherent control in nanoscale quantum transport.

In addition to promising applications, PI scars deserve further exploration in terms of classical-quantum correspondence away from the semiclassical limit. Instead of studying the probability density distribution of quantum states, the quantum fingerprints of chaos can be searched in the distribution of the energy levels -- along the methods widely applied in the field of quantum chaos.~\cite{Gutzwiller,Stockmann} In particular, a regular system can be characterized by an uncorrelated energy level sequence leading to a Poisson statistics of the eigenvalue spacings. On the other hand, chaotic behavior is related to the spectral statistics of random matrix ensembles. This kind of characterization is the essence of the famous Bohigas-Giannoni-Schmit conjecture (BGS).~\cite{BGS}

In this paper, we investigate the energy level statistics in perturbed two-dimensional oscillators which are known to show PI scarring. Upon the variation of the bumps and/or the magnetic field, scars evolve and the system undergoes transitions in the quantum eigenvalue spectrum. To describe spectral fluctuations, we employ the conventional eigenvalue statistics such as the spectral rigidity.~\cite{spectral_rigidity} This is supplemented with detrended fluctuation analysis (DFA)~\cite{DFA_intro1} -- an important tool in time-series analysis that can be applied also to characterize static complex systems.~\cite{DFA_intro2} Recently, DFA has been used in the classification of fractal magnetoconductance in chaotic quantum cavities.~\cite{kotimaki_rasanen_heller_PRE} We show that PI scarring is connected to transitions in the energy level statistics. In most cases, the scars can be detected from their surroundings in the mixed region between regular and chaotic eigenvalue statistics. An exception is the ``superintegrable'' 2D harmonic oscillator in a magnetic field~\cite{superintegrability_FD} that -- in the case of perturbation -- shows clear scars at magnetic fields that correspond to regular statistics. Finally, the connection between specific resonant energy eigenlevels and PI scarring is analyzed in detail by introducing the idea of the subspectrum of a scar. 

The paper is organized as follows. In Sec.~\ref{secsystem} we introduce the physical system and the model potential including the perturbation. In Sec.~\ref{secnumerics} we introduce the numerical scheme to solve the eigenvalue problem, and outline also the solutions of the corresponding classical system. In Sec.~\ref{secstatistics} we review the common schemes to assess quantum chaotic properties through eigenvalue statistics, and define estimators for mixed systems that combine regular and chaotic features. The results are divided to two parts: the relation of PI scarring and eigenvalue statistics in Sec.~\ref{secresults1} and the subspectrum analysis of particular eigenstates that contribute to PI scarring in Sec.~\ref{secresults2}. The main conclusions are discussed and the paper is summarized in Sec.~\ref{secsummary}.

\section{\label{secsystem}System}

All values and equations below are given in atomic units (a.u.). The Hamiltonian for a perturbed 2D quantum system has a form
\begin{equation}\label{quantum_Hamiltonian}
  H = \frac{1}{2}\big( -i\nabla + \mathbf{A} \big)^2 + V_{\text{ext}} + V_{\text{imp}}.
\end{equation}
The magnetic field $\mathbf{B}$ is oriented perpendicular to the 2D plane. It is included in the Hamiltonian through the vector potential $\mathbf{A}$. The external confinement potential is given by
\begin{displaymath}
V_{\text{ext}}(\mathbf{r}) = \frac{1}{2}\omega_0^2 \vert \mathbf{r} \vert^{n},
\end{displaymath}
where $n$ is an integer. We mainly focus on the cases of $n=5$ and $n=2$, where the latter corresponds to the harmonic oscillator that is widely used
as a model potential for semiconductor quantum dots.~\cite{reimann_manninen_RMP} We also briefly discuss alternative potentials, $n=8$ and a $\cosh$-type external potential.
In all cases, the factor of the confinement strength is set to $\omega_0=1$ for convenience. 

The perturbation $V_{\textrm{imp}}$ is modeled as a sum of Gaussian bumps:
\begin{displaymath}
V_{\text{imp}}(\mathbf{r}) = M \sum_{i} \exp\Bigg[ -\frac{(\mathbf{r} - \mathbf{r}_i)^2}{2\sigma^2}\Bigg].
\end{displaymath}
Here, $M$ and $2\sqrt{2} \sigma$ determine the amplitude and the full width at half maximum of the bumps, respectively. The bump width is comparable to the local wavelength of an eigenstate in the energy region where strong PI scarring is observed. The bumps are positioned randomly in the potential with a uniform mean density of two bumps per unit square. The PI scars appear in the energy range that corresponds to hundreds of bumps in the classically allowed region. However, even a single bump can produce a strong scar as demonstrated in Ref.~\onlinecite{controllability}, and -- in general -- PI scars are relatively common.\cite{Luukko} Here we do not quantitatively assess the dependence of the PI scars on the bump locations, i.e., we only consider one realization of random potential showing strong PI scarring. We focus on the general properties of the system, particularly the eigenspectrum and its subsets, as a function of $M$ and $\sigma$.

\section{\label{secnumerics}Numerical scheme}

We solve thousands of the eigenstates of the Hamiltonian of Eq.~(\ref{quantum_Hamiltonian}) by
employing the \texttt{itp2d} software package~\cite{itp2d}. The software utilizes the imaginary
time propagation method without a basis by orthonormalizing the states along the
propagation of initially random wave functions. The method is particularly suited for 2D systems with
perpendicular magnetic fields. This is due to the exact factorization of the exponential kinetic energy operator 
in a magnetic field.~\cite{itpmagn}

In the limit $M \rightarrow 0$ or $\sigma \rightarrow 0$, the Hamiltonian reduces to an unperturbed, circularly symmetric system, where the magnetic field lifts the
degeneracy of the states with opposite angular momenta. It is noteworthy that at denegeracies the individual (basis-free) solutions of \texttt{itp2d} may not be eigenstates
of the angular momentum. However, the common eigenstates of the Hamiltonian and the angular momentum operator can be formed as a linear combination of the 
degenerate numerical solutions. In the case of $n=2$ without perturbation, i.e., the 2D harmonic oscillator in a magnetic field, the Schr{\"o}dinger equation is 
analytically solvable. The energies correspond to the Fock-Darwin (FD) spectrum~\cite{FD}, and the corresponding eigenstates of the FD system can be expressed 
in the associate Laguerre polynomials (see, e.g., Ref.~\onlinecite{superintegrability_FD}).

In the corresponding {\em classical} system, the solution of the equations of motion can be found for a particle in an unperturbed symmetric $r^n$-potential without a magnetic field in 
standard texts on classical mechanics (see e.g., Ref.~\onlinecite{Goldstein}). 
The POs are associated with classical resonances where the oscillation frequencies of the radial
and the angular motion are commensurable. In this work, we use the 
notation $(v_{\theta}, v_{r})$ referring to a resonance, where the orbit circles the origin $v_{\theta}$ times in 
$v_{r}$ radial oscillations. If the potential $V(r)$ is a homogeneous 
function of the radius $r$, POs with different total energies are the same up to a scaling factor. 
A similar approach is possible in the case of a perpendicular 
homogeneous magnetic field.~\cite{ToBePublished} 
For a classical 2D HO in a magnetic field an analytical solution is available (see e.g., Ref.~\onlinecite{Kotkin}). In this case, resonances occur only at
\begin{equation}\label{magical_values}
B = \frac{v_r/v_{\theta} - 2}{ \sqrt{v_r/v_{\theta} - 1}}.
\end{equation}

Furthermore, we have studied the corresponding {\em perturbed} classical systems 
utilizing the \texttt{bill2d} software.~\cite{bill2d} The classical simulations have been conducted to confirm if the perturbation is sufficient to destroy classical 
long-term stability in the system.

\section{\label{secstatistics}Estimators of quantum chaos}

Next we describe the method to study the quantum solutions, in particular the energy 
eigenvalue spectrum resulting from the single-particle Schr\"odinger equation with
the Hamiltonian in Eq.~(\ref{quantum_Hamiltonian}). In the context of quantum chaos, several 
statistical measures have been developed to compare the energy spectra to the 
predictions given by the BGS, and to detect the quantum fingerprints of classical 
chaos.~\cite{Mehta}

First, the spectrum need to be unfolded to remove the non-universal contribution.~\cite{Eckhardt, Brody} 
In general, the average, smooth behavior of the density of states (DOS) is system-specific; 
the universal behavior is observed in the fluctuation around this mean. 
This fluctuating part of the initial spectrum can extracted by first 
determining the spectral staircase function
\begin{displaymath}
N(E) = \#\Big\{ E_m \, \Big \vert\, E_m \le E \Big\},
\end{displaymath}
and then separating it to the average and oscillating part:
\begin{displaymath}
N(E) = \bar{N}(E) + N_{\textrm{osc}}(E).
\end{displaymath}
Now, the unfolded energy spectrum is defined as $\varepsilon_m := \bar{N}(E_m)$. 
The unfolding can be a subtle process, see, e.g., Ref.~\onlinecite{unfolding}.

The simplest statistical measure of (quantum) chaos is the nearest-neighbor level spacing
(NNLS), i.e., the distribution $P(s)$ of distances $s$ between neighboring (unfolded) energy levels.
The NNLS displays short-range correlations in the spectrum. For a regular system, the energy levels 
are uncorrelated, and the NNLS distribution is Poissonian. In contrast, the chaotic limit is 
described by random matrix theory (RTM). Analytical expressions exist~\cite{Mehta} for 
Gaussian random-matrix ensembles, i.e., for the Gaussian orthogonal ensemble (GOE) and Gaussian 
unitary ensemble (GUE) with and without time-reversal symmetry, respectively. However,
here we employ the commonly applied RTM approximations, colloquially known as the Wigner surmise:\cite{Mehta}
\begin{displaymath}
P(s) \approx
\left\{ \begin{array}{ll}
\frac{\pi}{2}s e^{-\pi s^2/4}& \textrm{for GOE},\\
\\
\frac{32}{\pi^2}s^2 e^{-4 s^2/\pi} & \textrm{for GUE}.
\end{array} \right.
\end{displaymath}

The second statistical measure applied in this paper is the spectral rigidity $\Delta_3$ 
(see, e.g., Ref.~\onlinecite{Stockmann,Mehta}).  
It is defined as the integrated residual of the 
linear fittings, averaged over all intervals of size $L$:
\begin{displaymath}
\Delta_3(L) =  \Bigg \langle \min_{a,b} \int_{\varepsilon - L/2}^{\varepsilon + L/2} \Big[ n(\varepsilon) - a\varepsilon  -b \Big]^2\, \text{d}\varepsilon  \Bigg \rangle,
\end{displaymath}
where $n(\varepsilon)$ is the spectral staircase function of the unfolded energy spectrum.
Thus, the spectral rigidity measures the correlations in an energy window of size $L$. 
Here we employ the approximations for GOE and GUE distributions valid in the limit $L \gg 1$: 
\begin{displaymath}
\Delta_3(L) \approx
\left\{ \begin{array}{ll}
\frac{1}{\pi^2} \ln(L) - 0.007 & \textrm{for GOE},\\
\\
\frac{1}{2\pi^2} \ln(L) + 0.058 & \textrm{for GUE}.
\end{array} \right.
\end{displaymath}
Thus, the behavior in a chaotic system is logarithmic instead of the linear behavior 
in the regular case. However, the spectral rigidity should approach the Poisson result $L/15$, when $L \ll 1$. In addition, it has been shown~\cite{rigidity_saturation} that short periodic orbits cause a saturation of $\Delta_3$ to a finite value at large $L$.
The exact saturation value depends on the period (in time) of the orbit.

In many cases, a dynamical system is {\em mixed} in the sense that it shows
both chaotic and regular behavior. In a classical system, chaos sets in as the 
relative size of the non-integrable part of the Hamiltonian is increased according
to the Kolmogorov-Arnold-Moser theorem.~\cite{KAM_theory} In the quantum case,
we next review three ways to describe mixed behavior.

For the NNLS distributions, the Berry-Robnik mixing~\cite{BR_mixing} weights the
Poisson and GOE statistics with factors $0 \le q \le 1$ and $1-q$, respectively, 
leading to
\begin{displaymath}
\begin{split}
P(s) &= e^{qs} \Bigg[ q^2 \text{erf} \Big( \frac{\sqrt{\pi}}{2}(1-q)s \Big)\\ &+ \Big( 2q(1-q) + \frac{\pi}{2}(1-q)^3s \Big) e^{-\pi (1 -q)^2s^2/4} \Bigg]
\end{split}
\end{displaymath}
In the same spirit, we introduce a mixing parameter $Q$ for the spectral rigidity: 
\begin{displaymath}
\Delta_3^{\text{measured}}(L) = \Delta_3^{\text{Poisson}}(QL) + \Delta_3^{\text{GOE}/\text{GUE}}((1-Q)L).
\end{displaymath}
Here we determine the best fitting for an energy window $L$ 
instead of directly interpolating between the Poisson ($Q = 1$) and the GOE/GUE limit ($Q = 0$). In the results below, the energy window $L$ ranges from $0$ to $30$ which was found to be a sufficient maximum value. 

Another approach to analyze the spectral rigidity is to treat the fluctuations
in the quantum spectrum as a {\em time series}. It has been conjectured~\cite{Renalo,Faleiro}, 
that the power spectrum of spectral fluctuations follows a power law $f^{-\gamma}$, 
where $\gamma=1$ and $2$ correspond to the regular and chaotic limits, and a mixed system
has values $1<\gamma<2$. This definition is in line with the classical measure of chaos~\cite{pinknoise}
as demonstrated for the Robnik billiard~\cite{Robnik_billiard}.
In this work, we have applied a time series approach in terms of the DFA, which has an
explicit connection to the power spectrum analysis.
The DFA measures the scaling of spectral fluctuations as $\Delta_3(L) \sim L^{2\alpha}$,
where $\alpha\sim 0.5$ and $\alpha \sim 0$ correspond to the regular and chaotic limits as
described in Ref.~\onlinecite{DFA_and_spectrum}.

In the following, we apply the parameters $q$, $Q$, and $\alpha$ defined above to assess the level of 
chaoticity of quantum systems under consideration. However, we point out that there is no commensurable
definition of chaoticity resulting from energy level statistics.

\section{\label{secresults1}Energy level statistics}

\subsection{Zero magnetic field}

We begin by analyzing the $n=5$ case, i.e., the $r^5$-system. 
The system is interesting in terms of PI scarring~\cite{Luukko}
as its shortest non-trivial PO -- a five-point star (pentagram) --
is shorter than in the cases of $n = 1,3,4$ (the special case of $n=2$ is discussed below). 
It corresponds to a classical resonance of $(v_\theta,v_r)=(5,2)$. 

\begin{figure*}
\includegraphics[width=\linewidth, height= 0.25\linewidth]{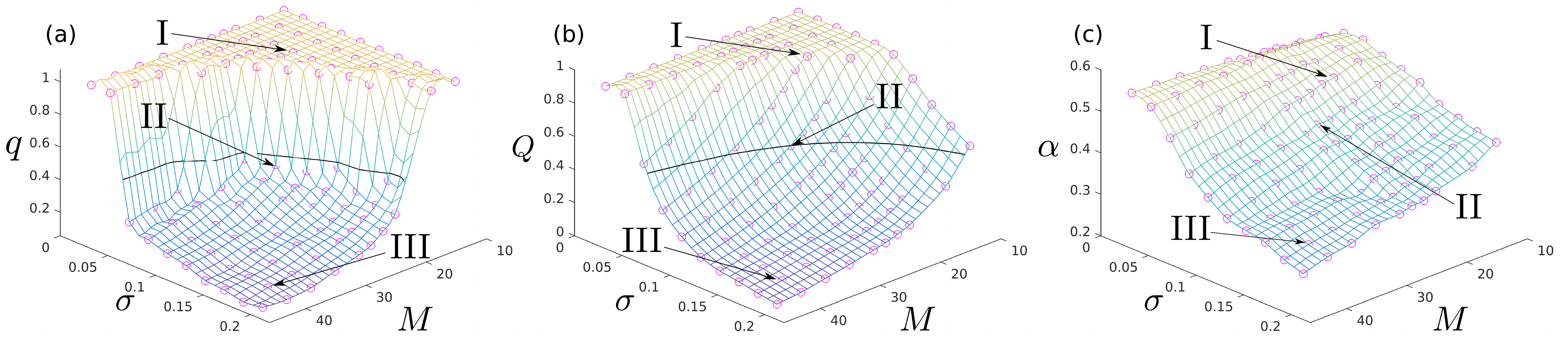}
\caption{Estimation parameters for chaoticity from eigenvalue level statistics in a perturbed $r^5$-type quantum system
as functions of the bump amplitude $M$ and width $\sigma$. 
(a) Berry-Robnik mixing parameter $q$ for the nearest-neighbor level spacing. 
(b) Mixing parameter $Q$ for spectral rigidity.
(c) Exponent $\alpha$ for spectral fluctuations according to the detrended fluctuation analysis.
The solid lines in (a) and (b) correspond $q = 0.5$ and $Q=0.5$, respectively. 
See text and Fig.~\ref{examples} for examples I, II, and III.\label{estimators}}
\end{figure*}

We solve 4000 energy eigenstates of Eq.~(\ref{quantum_Hamiltonian}) for each 
parameter combination of the bump amplitude $M = 12, \dots, 44$ and width $\sigma = 0.02, \dots, 0.20$.
From each spectrum, we compute the NNLS distribution and the spectral rigidity $\Delta_3$ with
the mixing parameters $q$ and $Q$, as well $\alpha$ from the DFA as defined in the previous section.
Figure~\ref{estimators} shows the results for these parameters as functions of $M$ and $\sigma$ as
open circles. The surfaces in the plots are constructed by applying cubic interpolation.

The NNLS mixing parameter $q$ in Fig.~\ref{estimators}(a) shows a relatively sharp transition
from regular behavior ($q$ close to one) towards the chaotic limit ($q=0$) 
as the amplitude or/and width is increased over a threshold value. The solid line depicts 
the value $q=0.5$. In the case of the spectral rigidity $Q$ in 
Fig.~\ref{estimators}(b), the transition occurs in the same parameter range, 
but it is smoother. However, $q$ and $Q$ are different at the leftmost parts of Figs.~\ref{estimators}(a) and (b), i.e., at large bump widths $\sigma$ and small amplitudes $M$. In this region, the potential is relatively smooth, but the bumps span over a large number of eigenstates, which affects the long-range correlations measured by $Q$. This effect is not captured by $q$ that measures NNLS mixing. Hence, the estimator $q$ remains close to the regular limit of $q=1$, whereas $Q < 1$.

The transition between regular and chaotic behavior is verified 
by $\alpha$ as seen in Fig.~\ref{estimators}(c). However, the transition is relatively smooth,
and the chaotic limit corresponding to $\alpha=0$ is not well captured, since the 
obtained range is $\alpha=0.3\ldots 0.5$. The transition is also very smooth compared
with $q$ and $Q$. In other words, the range $0\leq \alpha\lesssim 0.3$ is
missing in Fig.~\ref{estimators}(c), even though $q$ and $Q$ display full chaoticity at
large values of $M$ and $\sigma$. Therefore, the results indicate that -- particularly towards 
the chaotic limit -- $\alpha$ might not be an unambigious statistical measure compared to $q$ and $Q$.
Further studies are needed to utilize DFA more thoroughly in the assessment of quantum chaoticity.

\begin{figure*}
\includegraphics[width=\textwidth]{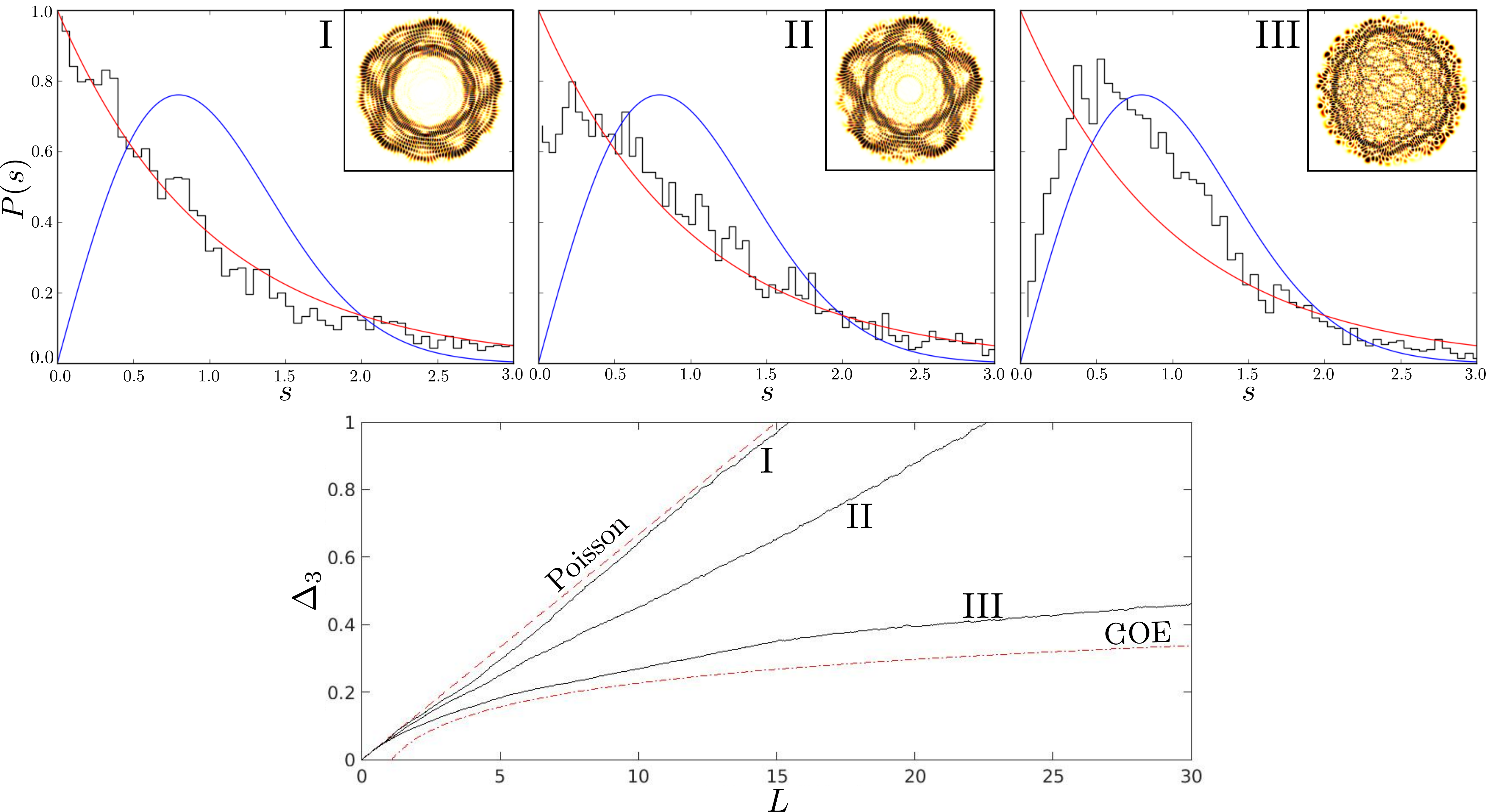}
\caption{Examples of eigenlevel statistics of the perturbed $r^5$-potential. The upper panel shows the nearest-neighbor level spacing distributions of Cases I, II, and III marked in Fig.~\ref{estimators}. The red and blue curves correspond the Poisson and GOE distributions, respectively. The insets show the probability density of the energy eigenstate $2791$, which is strongly scarred in Case II. In the lower panel, the spectral rigidities $\Delta_3$ of Cases I, II, and III are shown as a function of the energy window size $L$, along with the spectral rigidity of the Poisson and GOE limits.\label{examples}} 
\end{figure*}

Next, we consider in more detail three examples marked as I, II, and III in Fig.~\ref{estimators}.
The corresponding parameter combinations are $(M, \sigma)=(20, 0.08)$, $(26, 0.10)$, and $(40, 0.18)$, 
respectively. The upper panel of Fig.~\ref{examples} shows the NNLS distributions of these cases, 
and the lower panel shows the corresponding spectral rigidities. In Case I, the NNLS has the shape of the 
Possonian distribution, and the spectral rigidity follows the Poisson result. On the other hand, Case II 
located at the ``transition regime'', i.e., along the line $q=Q=0.5$ 
in Fig.~\ref{estimators} can be understood as a mixed quantum system. 
Its NNLS distribution and spectral rigidity have features of both Poisson and GOE statistics. Finally,
Case III located at the chaotic region in Fig.~\ref{estimators} shows GOE statistics in both NNLS and 
spectral rigidity in Fig.~\ref{examples}.

The insets in the upper panel of Fig.~\ref{examples} show the probability densities of state no. $2791$ (ordered in energy) at different levels of perturbation (I, II, III). PI scarring begins to occur in the parameter region near to the transition to chaos. In the chaotic regime, most of the scars observed in the eigenstates are faded into completely delocalized states. However, the distribution of Case III in Fig.~\ref{examples} has deviations from the GOE limit, which might be caused regular components such as remaining weak PI scars.  
Interestingly, as the amplitude and width of the bumps increase, the maximum visibility of a scar is observed around the middle of the transition in Figs.~\ref{estimators}(a) and (b). Nonetheless, the orientation of the scars does not change during their lifetime.

We want to address that PI scarring is not a rare occurrence (see also Ref.~\onlinecite{Luukko}). To provide statistical proof for the generality of our results, we have repeated the calculations for an ensemble of different random bump landscapes for three pairs of ($M$,$\sigma$) corresponding to Case I, II, and III, respectively. The size of the ensemble is 20, 40, and 20 bump configurations, respectively. For each mixing parameter $q$, $Q$, and $\alpha$, the average (avg.) and the standard deviation (SD) are determined. These results are summarized in Table~\ref{table_realizations}. The analysis confirm the validity of statistics for Cases I and II with relatively small SD. The largest SD occurs in Case II. In particular, the mixing estimator $q$ is rather sensitively on the considered bump configuration as the individual energy levels depend subtly on the locations of the bumps.
However, the estimator $Q$, as well as $\alpha$, measures long-range correlation in the spectra, and thus it is less sensitively on the difference between individual energy levels.
Nevertheless, Case II can be on average described to be located at transition region, where most visible PI scars are detected.
\begin{table}[h!]
\begin{center}
\begin{tabular}{c|cc|cc|cc}
Case & \multicolumn{2}{c}{$q$} & \multicolumn{2}{c}{$Q$} & \multicolumn{2}{c}{$\alpha$}\\
\hline
 & avg. & SD & avg. & SD & avg. & SD\\
\hline
\hline
I & $\sim 1$ & $\sim 0$ & $0.889$ & $0.037$ & $0.507$ & $0.009$\\
II & $0.593$ & $0.288$ & $0.466$ & $0.072$ & $0.439$ & $0.017$\\
III & $0.010$ & $0.015$ & $0.090$ & $0.028$ & $0.325$ & $0.017$\\
\end{tabular}
\caption{Average and standard deviation of the quantum estimators over many realizations of disorder: 20 random realizations of the bump locations for Cases I and III, and 40 for Case III. The parameters $(M, \sigma)$ are $(20, 0.08)$, $(26, 0.10)$ and $(40, 0.18)$, respectively, and they are same as used in Fig.~\ref{examples}.\label{table_realizations}}
\end{center}
\end{table}

\subsection{Non-zero magnetic field}

\begin{figure}
\begin{center}
\includegraphics[width = 1.0\linewidth]{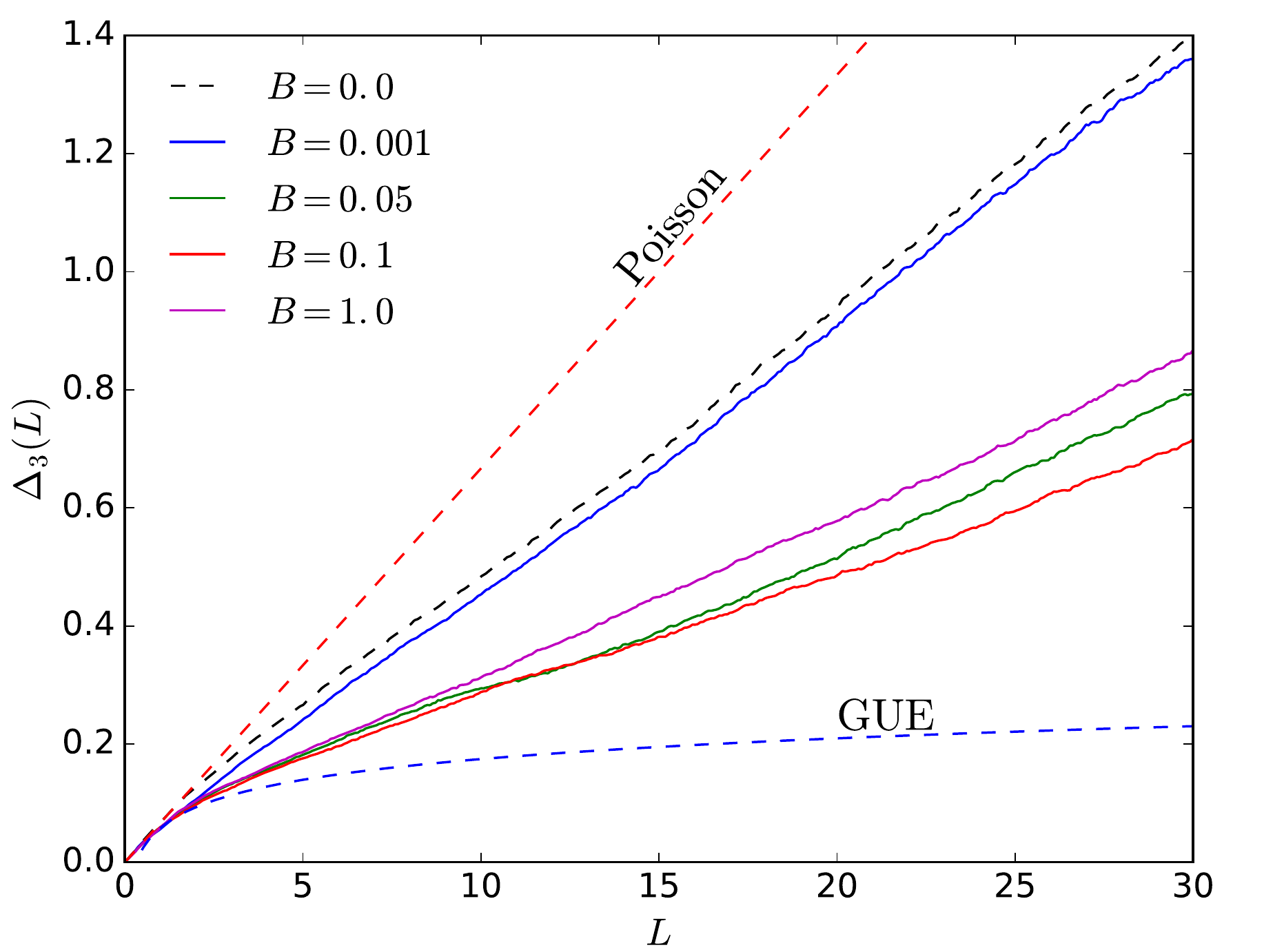}
\end{center}
\caption{Spectral rigidity of the perturbed $r^5$-potential with bumb amplitude $M = 4$ and width $\sigma = 0.1$ at different magnetic fields $B$. In addition to the computed spectral rigidities, the Poisson and the GUE limit are shown.  \label{effect_of_magnetic_field}} 
\end{figure}

Next we examine the effect of an external magnetic field on both the energy level statistics and scarring. Due to the breakdown of the time-reversal symmetry, the chaotic limit is described by the GUE statistics. In Fig.~\ref{effect_of_magnetic_field} we consider an $r^5$-system (as above) perturbed by bumps with amplitude $M = 24$ and $\sigma = 0.10$ under the influence of an external magnetic field. The parameters are chosen such that at $B=0$ strong PI scars are observed. As the strength of the field increases, the mixing estimators indicate that the system changes from regular behavior towards GUE statistics. However, the system does not reach the GUE limit even at high magnetic fields. Instead, the maximum ``chaoticity'' is achieved at $B \sim 0.1$. At higher fields, the system starts to move back towards the Poisson limit. This can be understood in a way that a high magnetic field effectively regularizes the system due to strong Lorentz forces that enhance regularity.

As the magnetic field is increased, the relative number of PI scars decreases, although they are present even at strong fields ($B \sim 1$). The observed pentagram scars have a preferred orientation due to the magnetic field. We confirmed numerically the existence of classical $(5,2)$ orbits in a pure system with a magnetic field.~\cite{ToBePublished}

Finally, we focus on a harmonic oscillator ($r^2$ system) in an external magnetic field, which is an exceptional system regarding the PI scarring. In Ref.~\onlinecite{controllability}, it was shown that strong quantum scars occur at specific magnetic field values that determine the geometry of the scar through the corresponding classical resonance $(v_{\theta}, v_r)$. 

\begin{figure*}
\includegraphics[width=\textwidth]{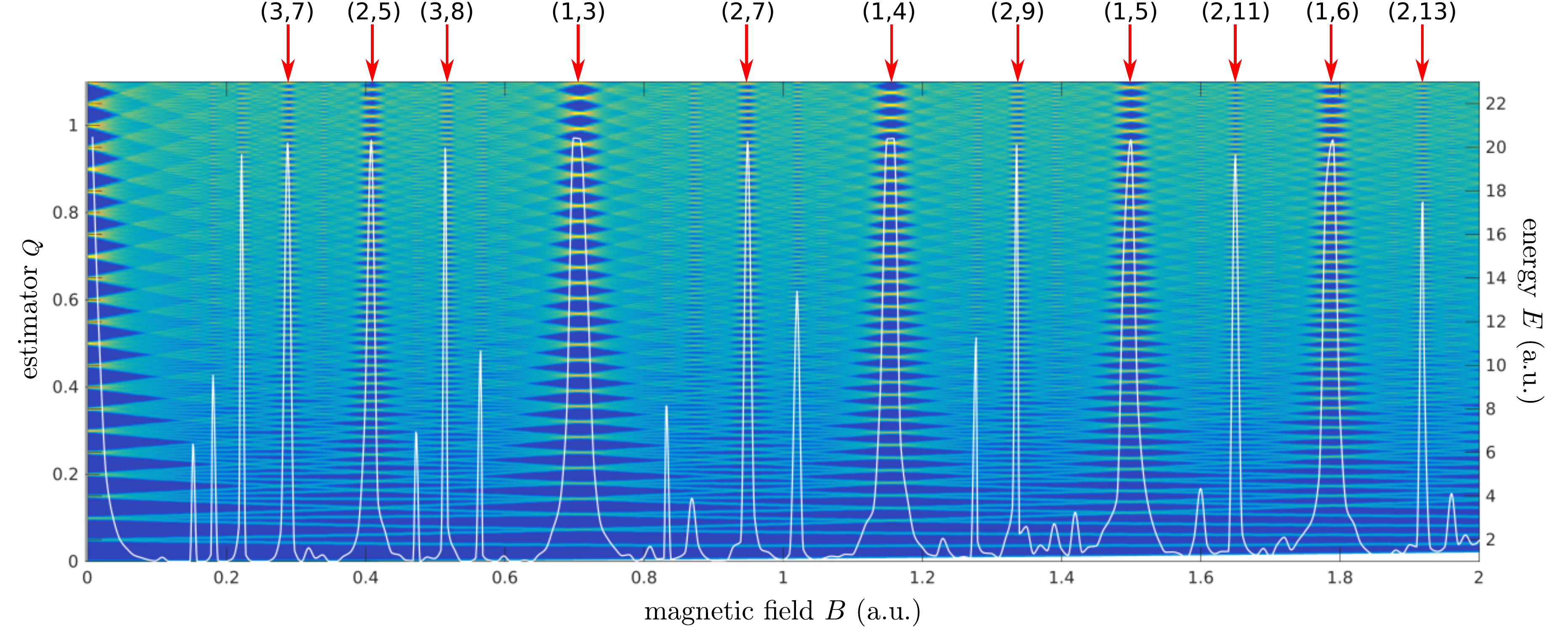}
\caption{Density of states (in arbitrary units) of the unperturbed harmonic oscillator as a function of the magnetic field $B$, and energy $E$. The arrows indicate the resonances $(v_{\theta}, v_r)$ corresponding to a substantial abundance of scarred eigenstates when the oscillator potential is {\em perturbed} by bumps~\cite{controllability}. The white curve shows the mixing estimator $Q$ (Poisson vs. GUE) computed from the spectral rigidity in a perturbed system with $(M,\sigma)=(4,0.1)$ as a function of the magnetic field. \label{q_value_and_dos}} 
\end{figure*}

Figure~\ref{q_value_and_dos} shows the density of states (DOS) of the analytic FD spectrum computed as a sum of the states in a Gaussian energy window of $0.001$ a.u. The white curve on top of the DOS shows in the spectral rigidity mixing estimator $Q$ (Poisson vs. GUE) calculated from 4000 energy eigenstates of the {\em perturbed} oscillator with bump amplitude $M = 4$ and width $\sigma = 0.1$. The eigenlevel statistics has been calculated separately for each value of $B=0\ldots 2$, with a resolution of $\Delta B = 0.01$. The perturbation causes strong PI scarring in the energy range $E = 50, \dots, 100$. Some of the $(v_{\theta}, v_r)$ resonances are marked by arrows.~\cite{controllability}

As seen in Fig.~\ref{q_value_and_dos} (white curve), the energy level statistics of the perturbed system changes rapidly from Poisson-like $(Q=1)$ towards GUE $(Q\sim 0)$ as $B$ is increased from zero. At specific values of $B$, however, corresponding to the resonances at near-degeneracies, significant deviations from GUE statistics are observed. Interestingly, particularly strong scarring is seen as $Q$ is closer to the Poisson limit. Vice versa, when scars are not visible, the system is close to chaotic limit described by the GUE statistics. In this sense, the PI scars correspond to order in a perturbed HO. This is also seen in Fig.~\ref{q_value_and_dos} as high degeneracies of the energy states at resonances.

\section{\label{secresults2}Subspectrum analysis}

Besides analyzing the statistics of the full spectrum and their correspondence to the scarring effect,
 it is worthwhile to examine how individual states contribute to PI scars. We begin by briefly considering
the mechanism behind PI scarring in the framework of quantum perturbation theory as explained 
in Ref.~\onlinecite{Luukko}.

First, the unperturbed circularly symmetric system is separable leading to special near-degeneracies which 
are connected to classical resonances. The eigenstates are labeled by two quantum numbers $(r, l)$, 
corresponding to radial and angular motion, respectively. At $B=0$, the opposite angular momentum states 
$(r, \pm l)$ are exactly degenerate. Additionally, there are near-degeneracies related to the classical POs.
Based on the Bohr-Sommerfield quantization, if a state $(r, l)$ is nearby in action to a periodic orbit 
corresponding to a resonance of the oscillation frequencies $(v_{\theta}, v_{r})$, the states 
$(r + k v_{\theta}, l - k v_{r})$ with small $k \in \mathbb{N}$ are nearby in energy. These states are referred to 
a ``resonant set'' of the unperturbed basis states. Because of the relationship of the resonant states to the 
classical resonances, some linear combinations in the resonant subspace will trace out the path of the 
classical periodic orbit $(v_{\theta}, v_{r})$.

Secondly, a sufficiently small perturbation leads to the localization of the eigenstates mostly at the 
near-degenerate part of the resonant subspace. Due to the localized nature of the bumps, the system prefers 
scarred linear combinations, which effectively extermize the perturbation. 

Next, we present the concept of the subspectrum of a scar. A scarred state $\vert \psi \rangle$ of a perturbed 
quantum system is expanded in the basis of the corresponding unperturbed 
states $\vert \phi_m \rangle$ ($m \in \mathbb{N}$) in the following way,
\begin{displaymath}
\vert \psi \rangle = \sum_m c_m \vert \phi_m \rangle,
\end{displaymath}
where $c_m \in \mathbb{C}$. We refer to the set $\{\vert c_m \vert^2\}$ as a subspectrum of a scar. The values $\vert c_m \vert^2 \in \mathbb{R}$ are restricted to interval $[0,1]$ for all $m \in \mathbb{N}$, since the state $\vert \psi \rangle$ and the basis states $\vert \phi_m \rangle$ are orthonormalized. Therefore, the subspectrum describes the relative weights of the unperturbed states in the construction of a PI scar.

\begin{figure*}
\includegraphics[width=\textwidth]{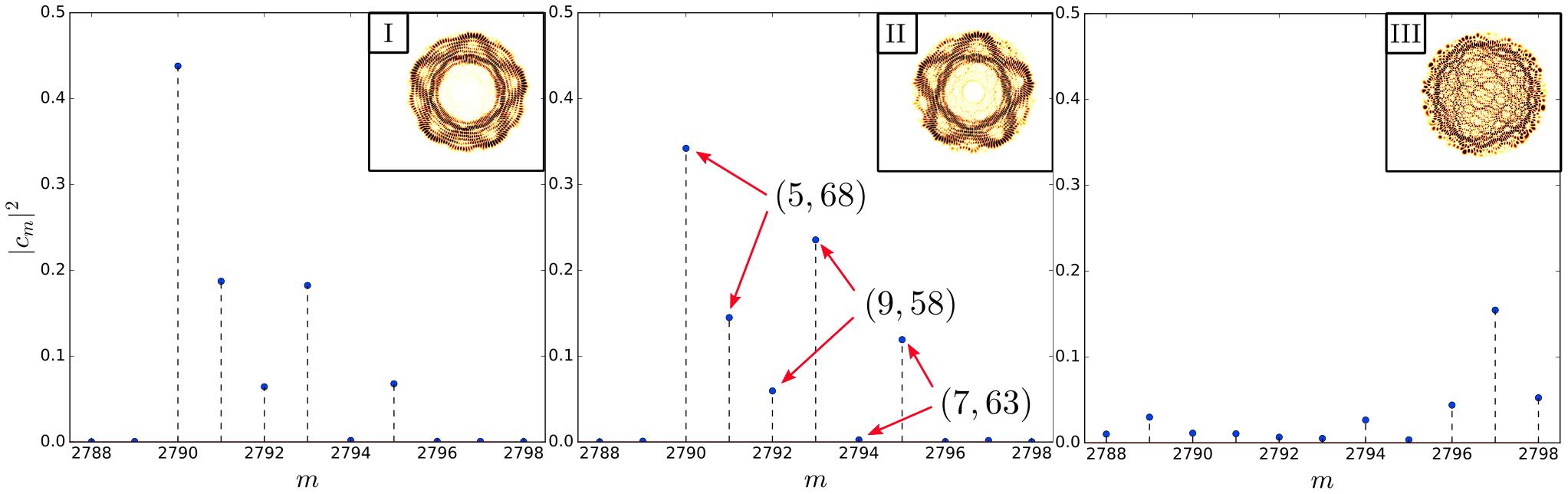}
\caption{Example of a state (no. 2791) in a perturbed system in the basis of unperturbed states (see text). 
The cases I, II, and III correspond to those considered in Figs. 1 and 2. In the middle panel, the 
quantum numbers $(r, \vert l \vert)$ are shown for the essential unperturbed states that form the
scar (inset). \label{subspectrum}}
\end{figure*}

Figure~\ref{subspectrum} shows the subspectrum of the energy eigenstate 2791 in the perturbed quantum system
corresponding to Cases I, II, and III -- the same examples as those considered above in Figs.~\ref{estimators} and~\ref{examples}. 
The basis of the unperturbed energy eigenstates is the same in all cases. The insets show the  
probability densities of the states; the same as those in Fig. \ref{examples}. In the middle panel of 
Fig.~\ref{subspectrum}, the radial and angular quantum numbers $(r, \vert l \vert)$ are marked 
for relevant unperturbed states. 

According to the two leftmost panels in Fig.~\ref{subspectrum}, only five states states have a clear 
contribution to the subspectrum of the state in the Cases I and II. The states belong to a resonant
set $(r + 2k, l - 5k)$, where $k \in \mathbb{N}$. The contribution is most evenly spread in the resonant
set in Case II, where the interference pattern traces out the shape of the classical PO, i.e., a pentagram, and PI
scarring is particularly strong. In Case I with a weaker perturbation, the unperturbed state $m = 2790$ 
dominates the subspectrum. Consequently, the perturbed state resembles more a circularly symmetric state 
of the unperturbed system than a PI scar.

In the chaotic region, corresponding to Case III and the rightmost panel of Fig.~\ref{subspectrum},
the perturbation is strong enough to couple many unperturbed states. Therefore, the probability density 
distribution seems ``chaotic'', reflecting the nature of ergodicity of the underlining classically chaotic 
system. However, classical simulations~\cite{ToBePublished} reveal that the perturbation in Case II is already 
sufficient to destroy any long-time stability. The remaining structures in the otherwise chaotic Poincaré 
surface of section are vanishingly small compared to $\hbar = 1$ (see also Ref.~\onlinecite{Luukko}).

\section{\label{secsummary} Discussion and summary}

Our results generally show that the most visible scars are detected as the system undergoes a transition 
from the regular to chaotic region. The perturbation is then sufficient to couple the relevant unperturbed states 
but it is not too strong to destroy the near-degenerate structure of the resonant sets. 
However, the details depend on the given potential, in particular, which values of $(M,\sigma)$ can
be considered as a {\em sufficient} perturbation. In addition to the $r^5$ and $r^2$ potentials considered
in detail above, we have examined a $r^8$ potential as well as a non-homogeneous yet circularly
symmetric $\cosh$-type potential. In both cases, we observe similar behavior in energy level statistics compared
to the $r^5$ case: an increase in the amplitude and the width of the bumps result in a transform from the 
Poisson to the GOE statistics. The most visible PI scarring arises at the intermediate region. 

The external magnetic field has a similar qualitative effect on the eigenvalue statistics 
in different potentials. With a fixed perturbation, an increasing $B$ eventually begins to dominate 
the properties of the system. In the classical picture, the Lorentz force begins to overcome the perturbation 
caused by the bumps in the potential. On the other hand in the quantum system, energy levels start to 
condensate into Landau levels. This results in increasing regularity in the estimators of energy level statistics.

The special characteristics of a perturbed $r^2$ system (HO) in a magnetic field arise from its superintegrability. 
A quantum system is called superintegrable~\cite{superintegrability} if there is a maximal set of independent 
symmetry operators, i.e., additional symmetry operators (not necessary commuting) exist compared to an 
integrable system. The two limits of the unperturbed FD system at $B=0$ and $B\rightarrow\infty$ (Landau system)
are known to be superintegrable, and a general FD system is superintegrable at specific magnetic field values
as shown in Ref.~\onlinecite{superintegrability_FD}. These values correspond to Eq.~(\ref{magical_values}),
and their neighborhood upon sufficient perturbation shows PI scarring.~\cite{controllability} Furthermore,
as shown above, the superintegrability leads to strong peaks of regularity in estimator $Q$ in 
Fig.~\ref{q_value_and_dos}. It is noteworthy that the PI scars resemble the coherent states of the 
unperturbed FD system presented in Ref.~\onlinecite{superintegrability_FD}.

In addition, we want to point out that the PI scarring cannot be explained by dynamical localization.~\cite{dynamical_localization_I, dynamical_localization_II} Although present in systems studied here, dynamical localization corresponds to localization in angular momentum space, whereas the scars are localized in position space. Furthermore, dynamical localization does not explain the preferred orientations of the PI scars (see Ref.~\onlinecite{Luukko}).

In general, the present work shows how the recently discovered form of quantum scarring -- PI scarring --  is related to eigenvalue statistics. Our results assist to develop the theory of PI quantum scarring further as well as to understand the connection between PI scars and related phenomena such as conventional scarring~\cite{Heller, Kaplan} and branched flow~\cite{branched_flow}. Moreover, for classical billiards it has been shown that soft boundaries can bring chaos.~\cite{classical_chaos_softness} In addition, an external magnetic field can cause chaotic behavior.~\cite{classical_chaos_magnetic field} It has been hypothesized~\cite{fractal_conductance_theory} that the softness of the wall combined with a magnetic field causes fractal behavior of the magnetocondutance observed in semiconductor quantum cavities.~\cite{fractal_conductance_experiment} Therefore, we provide a statistical study on this kind of distorted, realistically bound (soft boundary) quantum cavities including the effect of an external magnetic field. Furthermore, as previous studies have demonstrated~\cite{quantum_control}, classical chaos can also be exploited to control quantum transport. Likewise, our research aims to the development of a quantum control scheme: it paves way towards ``scartonics'', where PI scars are utilized to coherently control conductance in nanoscale quantum systems (see Refs.~\onlinecite{Luukko, controllability}).

To summarize, we have studied the eigenvalue statistics and several estimators for quantum chaos in the
case of two-dimensional quantum wells perturbed by randomized bumps in the potential. We have focused on
the connection between the statistics and the formation of perturbation-induced quantum scars, i.e.,
eigenstates that resemble classical periodic orbits of the corresponding unperturbed system. 
We have shown that the system undergoes a smooth transition from regular behavior to chaos as the
perturbation is increased. The used estimators for chaoticity, i.e., the Berry-Robnik and spectral rigidity
mixing parameters as well as the exponent of detrended fluctuation analysis are in a qualitative
agreement in the transition. The perturbation-induced scarring is strong in the transition regime
of mixed eigenvalue statistics. The results between different potentials, and in the presence of
an external magnetic field are consistent. An exception is the superintegrable Fock-Darwin system,
where the regime of scarring is characterized by Poisson-like behavior in the chaos estimators.
We have also demonstrated in detail the scar formation and composition at the subspectral level.

The present results show that systems with perturbation-induced quantum scars are (mostly) {\em mixed} 
in terms of the conventional statistical measures of quantum chaos, even though the corresponding classical systems
are highly chaotic as already analyzed in a previous work.~\cite{Luukko}
The relationship between the energy level statistics and the PI scarring is shown to stem from the resonant sets, i.e., particular near-degeneracies, in the corresponding unperturbed system and the the local nature of the perturbations.
However, further research is required
both to further rationalize quantum chaotic estimators in mixed systems and to exploit scarring in quantum technology.

\section*{Acknowledgments}

This work was supported by the Academy of Finland. We also acknowledge 
CSC -- Finnish IT Center for Science -- for computational resources. 
We are grateful to Prof. Eric Heller and Prof. Lev Kaplan for useful discussions.
We also thank Janne Solanp{\"a}{\"a}, and Matti Molkkari for assisting in classical 
phase space analysis, and the DFA analysis, respectively.

\end{document}